\documentstyle[12pt]{article}
\renewcommand{\thefootnote}{\fnsymbol{footnote}}

\newlength{\extraspace}
\newlength{\extraspaces}
\newcommand{\be}{\begin{equation}
\addtolength{\abovedisplayskip}{\extraspaces}
\addtolength{\belowdisplayskip}{\extraspaces}
\addtolength{\abovedisplayshortskip}{\extraspace}
\addtolength{\belowdisplayshortskip}{\extraspace}}
\newcommand{\ee}{\end{equation}}
\newcommand{\ba}{\begin{eqnarray}
\addtolength{\abovedisplayskip}{\extraspaces}
\addtolength{\belowdisplayskip}{\extraspaces}
\addtolength{\abovedisplayshortskip}{\extraspace}
\addtolength{\belowdisplayshortskip}{\extraspace}}
\newcommand{\ea}{\end{eqnarray}}
\newcommand{\newsection}[1]{
\vspace{7mm}
\pagebreak[3]
\addtocounter{section}{1}
\setcounter{equation}{0}
\setcounter{subsection}{0}

{\large {\bf \thesection. #1}}
\nopagebreak
\medskip
\nopagebreak
\hspace{3mm}}
\newcommand{\nonu}{\nonumber \\[.5mm]}
\newcommand{\A}{&\!\!\!}

\begin{document}
\addtolength{\baselineskip}{3.0mm}
\begin{flushright}
SIT-HEP-00/10 \\
October, 2000
\end{flushright}
\vspace{.1cm}

\begin{center}
{\large{\bf{$N = 3$ chiral supergravity 

            compatible with the reality condition 

and 

higher $N$ chiral Lagrangian density}}} 
\\[10mm]
{\large Motomu Tsuda}
\footnote{e-mail: tsuda@sit.ac.jp} 
\\
Laboratory of Physics, Saitama Institute of Technology \\
Okabe-machi, Saitama 369-0293, Japan 
\\[10mm]

{\bf Abstract}\\[7mm]
{\parbox{13cm}{\hspace{5mm} 
We obtain $N = 3$ chiral supergravity (SUGRA) 
compatible with the reality condition 
by applying the prescription of constructing 
the chiral Lagrangian density from the usual SUGRA. 
The $N = 3$ chiral Lagrangian density in first-order 
form, which leads to the Ashtekar's canonical 
formulation, is determined so that it reproduces 
the second-order Lagrangian density of the usual SUGRA 
especially by adding appropriate four-fermion 
contact terms. We show that the four-fermion 
contact terms added in the first-order chiral 
Lagrangian density are the non-minimal terms 
required from the invariance under 
first-order supersymmetry transformations. 
We also discuss the case of higher $N$ theories, 
especially for $N = 4$ and $N = 8$.}} 
{\parbox{13cm}{\hspace{5mm} 
PACS number(s): 04.65.+e}} 
\end{center}
\vfill

\newpage

\renewcommand{\thefootnote}{\arabic{footnote}}
\setcounter{section}{0}
\setcounter{equation}{0}
\setcounter{footnote}{0}
\newsection{Introduction}

The supersymmetric extension of the Ashtekar's 
canonical gravity \cite{AA,AA2} has been developed 
since the first construction of $N = 1$ chiral 
\footnote{
The `chiral' means in this paper 
that only right-handed (or left-handed) 
spinor fields are coupled to the spin connection 
in the kinetic term of spinor fields.}
supergravity (SUGRA) \cite{Jac}. 
In particular, the extended chiral SUGRA 
was constructed in the context of the two-form 
gravity \cite{Pl,CDJ} for $N = 2$ \cite{KS,Ez} 
and $N = 3,\ 4$ theories \cite{KN}, and was also 
constructed by closely following the method of 
the usual SUGRA for $N = 2$ theory \cite{TS1}. 
Furthermore the canonical formulation of SUGRA 
in terms of the Ashtekar variable was explicitly 
derived up to $N = 2$ theory from 
the method of the two-form SUGRA \cite{KS,Ez} 
and also from that of the usual SUGRA \cite{TS2}. 
However, for $N \ge 3$ chiral SUGRA, 
the straightforward derivation of the canonical 
formulation in terms of the Ashtekar variable 
has not yet been done in the literature so far. 

In this paper we construct $N = 3$ chiral SUGRA 
compatible with the reality condition, 
which is the lowest $N$ theory involving a spin-1/2 field 
in addition to spin-2 (gravitational), 
spin-3/2 and spin-1 fields, 
by closely following the method of the usual SUGRA 
as a preliminary to derive the canonical formulation 
of $N = 3$ SUGRA in terms of the Ashtekar variable. 
Furthermore, we discuss the construction 
of higher $N$ chiral SUGRA, in particular, 
the construction of the chiral Lagrangian density 
for $N = 4$ and $N = 8$ theories. 

When we construct the chiral SUGRA, 
we assume at first that the tetrad is complex 
and construct such a chiral Lagrangian as analytic 
in complex field variables as briefly mentioned 
in \cite{Jac}. 
This means that right- and left-handed SUSY 
transformations introduced in the chiral SUGRA are 
independent of each other even in the second-order 
formalism. 
This fact makes it more transparent to confirm 
the SUSY invariance, particularly the right-handed one. 
Once we construct the chiral Lagrangian density, 
we impose the reality condition. 
\footnote{
After imposing the reality condition, 
the phase space in the canonical formulation 
consists of the real triad 
and the complex connection 
as discussed, for example, 
in \cite{AA,AA2,FK,TUV} for general relativity, 
and in \cite{Jac,MM} for ($N = 1$) chiral SUGRA. 
This phase space leads to non-Hermitian operator 
with respect to the connection variable 
in quantizing the theory 
(see, for example, Refs. \cite{FK,MM}).}

In order to construct the chiral Lagrangian density 
{\it in first-order form}, we apply the prescription 
proposed in the case of $N = 2$ chiral SUGRA \cite{TS1}. 
Firstly the chiral Lagrangian density 
in the second-order formalism, 
${\cal L}^{(+)}[{\rm second\ order}]$, 
with a (complex) tetrad is obtained 
from the Lagrangian density of the usual SUGRA, 
${\cal L}_{{\rm usual\ SUGRA}}[{\rm second\ order}]$, 
by complexifying spinor fields (spin-3/2 and -1/2 fields) 
as follows: 
\begin{enumerate}
\def\labelenumi{(\theenumi)}
\def\theenumi{\alph{enumi}}
\item 
Replace Rarita-Schwinger fields 
\footnote{
Rarita-Schwinger fields denoted by $\psi^I_{\mu}$ 
and $\tilde \psi^I_{\mu}$, 
and spin-1/2 fields 
denoted by $\chi$ (or $\chi^I$, $\chi^{IJK}$) 
and $\tilde \chi$ 
(or $\tilde \chi^I$, $\tilde \chi^{IJK}$) 
represent Majorana spinors. 
Throughout this paper capital letters 
$I, J, \dots$ denote 
the number of Rarita-Schwinger fields, 
and we shall follow the notation 
and convention of Ref. \cite{TS1}.}
$\psi^I_{\mu}$ and their conjugates 
$\overline \psi^I_{\mu}$ with 
\ba
\A \A \psi^I_{\mu} \rightarrow 
      \psi^I_{R \mu} + \tilde \psi^I_{L \mu}, 
      \nonu
\A \A \overline \psi^I_{\mu} \rightarrow 
      \overline \psi^I_{L \mu} 
      + \overline{\tilde \psi}^I_{R \mu}, 
\ea
by using two independent sets of Rarita-Scwinger 
fields $\psi^I_{\mu}$ and $\tilde \psi^I_{\mu}$. 

\item 
Rewrite the kinetic term of 
$e \epsilon^{\mu \nu \rho \sigma} \overline \psi^I_{L \mu} 
\gamma_{\rho} \nabla_{\sigma} \tilde \psi^I_{L \nu}$ 
into $e \epsilon^{\mu \nu \rho \sigma} 
\overline{\tilde \psi}^I_{R \mu} 
\gamma_{\rho} \nabla_{\sigma} \psi^I_{R \nu}$ 
plus a total derivative by partial integration, 
where $\nabla_{\sigma}$ denotes the ordinary covariant 
derivative in general relativity. 

\item
Apply the prescription (a) and (b) to 
spin-1/2 fields. 

\end{enumerate}
Then the chiral Lagrangian density in first-order form, 
${\cal L}^{(+)}$, is determined 
by the following prescription: 
\begin{description}
\item{(d)} Replace the $\nabla_{\sigma}$ 
to the $D^{(+)}_{\sigma}$ which is defined 
in Eq. (\ref{D+}) later. 

\item{(e)} Add appropriate four-fermion contact terms 
so that the ${\cal L}^{(+)}$ reproduce the 

${\cal L}^{(+)}[{\rm second\ order}]$. 

\end{description}
The four-fermion contact terms 
added in ${\cal L}^{(+)}$ by means of the prescription (e) 
are also required from the invariance under first-order 
supersymmetry (SUSY) transformations 
up to $N = 3$ chiral SUGRA as will be explained later: 
We expect that the same result will be the case 
for higher $N$. 
The ${\cal L}^{(+)}$ in terms of the real spin contents 
is obtained by imposing the reality condition 
(e.g., $\overline{e^i_{\mu}} = e^i_{\mu}$ for the tetrad, 
and $\tilde \psi^I_{\mu} = \psi^I_{\mu}$ 
for Rarita-Schwinger fields). 

In the case of $N = 2$ chiral SUGRA \cite{TS1} 
{\it under the reality condition}, 
the chiral Lagrangian density in first-order form 
constructed by the above prescription (a)-(e) 
differs from the first-order Lagrangian density 
of the usual $N = 2$ SUGRA by 
\ba
\A \A ({\cal L}^{(+)}_{N = 2} 
- {\cal L}_{N = 2 {\rm\ usual\ SUGRA}}) \ 
[{\rm first\ order}] 
\nonu
\A \A = - {i \over {8 \kappa^2}} 
e \ \epsilon^{\mu \nu \rho \sigma} 
(T_{\lambda \mu \nu} + i \kappa^2 
\overline \psi^I_{\mu} \gamma_{\lambda} 
\psi^I_{\nu}) \ T{^{\lambda}}_{\rho \sigma} 
+ {i \over 8} \kappa^2 
e \ \epsilon^{\mu \nu \rho \sigma} 
(\overline \psi^I_{L \mu} \psi^J_{R \nu}) 
\overline \psi^K_{R \rho} \psi^L_{L \sigma} 
\epsilon^{IJ} \epsilon^{KL} 
\nonu
\A \A \ \ \ 
- {i \over {4 \kappa^2}} \partial_{\mu} 
\{ e \ \epsilon^{\mu \nu \rho \sigma} 
(T_{\nu \rho \sigma} + i \kappa^2 
\overline \psi^I_{\rho} \gamma_{\nu} 
\psi^I_{\sigma}) \}, 
\label{LN2}
\ea
where $T_{\lambda \mu \nu}$ stands for the torsion tensor. 
The last imaginary boundary term corresponds to 
a certain Chern-Simons boundary term 
given by Mac\'\i as \cite{Mac} and Mielke et al. \cite{MMT} 
as a generating function of the canonical transformation. 
However, {\it in the first-order formalism}, 
the second four-fermion contact term added 
by the prescription (e) does not appear 
in $N = 1$ chiral SUGRA \cite{Jac,Mac,MMT}. 
Indeed, we showed that this new second term is 
the non-minimal one required from the invariance 
under first-order SUSY transformations \cite{TS1}. 
In the second-order formalism, the first term 
does not vanish by itself in contrast with 
the $N = 1$ theory, but cancels with 
the second four-fermion contact term 
by using a Fierz transformation. 
In $N = 3$ chiral SUGRA, an additional four-fermion 
contact term quadratic with respect to both spin-3/2 
and -1/2 fields is also required in the chiral 
Lagrangian density as explained in the next section. 

This paper is organized as follows. In Sec. 2 we construct 
$N = 3$ chiral SUGRA compatible with the reality condition 
by applying the above prescription (a)-(e). 
The invariance of the field equation for vector fields 
under duality transformations in $N = 3$ chiral SUGRA 
is shown in Sec. 3. In Sec. 4 we discuss the case of 
higher $N$ theories, especially for $N = 4$ and $N = 8$. 
The conclusion is given in Sec. 5.

\newsection{$N = 3$ chiral SUGRA compatible 
with the reality condition}

Let us construct $N = 3$ chiral SUGRA 
by means of the prescription (a)-(e) 
explained in Introduction. 
The usual $N = 3$ SUGRA \cite{Fr,FSZ} 
has spin contents $(2, {3 \over 2}, 
{3 \over 2}, {3 \over 2}, 1, 1, 1$, ${1 \over 2})$. 
Corresponding to these spin contents, 
the independent variables in $N = 3$ chiral SUGRA 
are a (complex) tetrad $e^i_{\mu}$, 
two independent sets of Rarita-Schwinger 
fields ($\psi^I_{\mu}, \tilde \psi^I_{\mu}$) 
($I = 1, 2, 3$), two independent spin-1/2 fields 
($\chi, \tilde \chi$), 
(complex) vector fields $A^I_{\mu}$, 
in addition to the self-dual connection 
$A^{(+)}_{ij \mu}$ which is also treated as 
one of the independent variables 
in the first-order formalism. 
Then the $N = 3$ chiral Lagrangian density 
in first-order form is given by 
\ba
{\cal L}^{(+)}_{N = 3} 
\A = \A - {i \over {2 \kappa^2}} 
        e \ \epsilon^{\mu \nu \rho \sigma} 
        e^i_{\mu} e^j_{\nu} R^{(+)}_{ij \rho \sigma} 
      - e \ \epsilon^{\mu \nu \rho \sigma} 
        \overline{\tilde \psi}^I_{R \mu} \gamma_{\rho} 
        D^{(+)}_{\sigma} \psi^I_{R \nu} 
\nonu
\A \A + i e \ \overline{\tilde \chi}_R 
        \gamma^{\mu} 
        D^{(+)}_{\mu} \chi_R 
      - {e \over 4} (F^I_{\mu \nu})^2 
\nonu
\A \A + {\kappa \over{2 \sqrt{2}}} e \ \{ 
        (F^{(-)I \mu \nu} + \hat F^{(-)I \mu \nu}) 
        \overline \psi^J_{L \mu} \psi^K_{R \nu} 
\nonu
\A \A \hspace{1.5cm}
      + (F^{(+)I \mu \nu} + \hat F^{(+)I \mu \nu}) 
        \overline{\tilde \psi}^J_{R \mu} 
        \tilde \psi^K_{L \nu} \} \epsilon^{IJK} 
\nonu
\A \A - {i \over 2} \kappa \ e 
        \left\{ \hat F^I_{\mu \nu} 
        - {\kappa \over 2} (\overline \psi^I_{L \mu} 
        \gamma_{\nu} \tilde \chi_L 
        + \overline{\tilde \psi}^I_{R \mu} 
        \gamma_{\nu} \chi_R) \right\} 
        (\overline \psi^I_{L \lambda} 
        S^{\mu \nu} \gamma^{\lambda} \tilde \chi_L 
        + \overline{\tilde \psi}^I_{R \lambda} 
        S^{\mu \nu} \gamma^{\lambda} \chi_R) 
\nonu
\A \A + {i \over 8} \kappa^2 
        e \ \epsilon^{\mu \nu \rho \sigma} 
        (\overline \psi^J_{L \mu} \psi^K_{R \nu}) 
        \overline{\tilde \psi}^L_{R \rho} 
        \tilde \psi^M_{L \sigma} 
        \epsilon^{IJK} \epsilon^{ILM} 
\nonu
\A \A + {\kappa^2 \over 2} e \ 
        (\overline{\tilde \psi}^I_{R \mu} 
        \gamma^{[\mu} \psi^I_{R \nu}) \ 
        \overline{\tilde \chi}_R \gamma^{\nu]} 
        \chi_R, 
\label{LN3}
\ea
which is globally O(3) invariant. 
Here $e := {\rm det}(e^i_{\mu})$, 
$\epsilon^{IJK}$ denotes a totally antisymmetric tensor, 
and $F{^{(\pm)I}}_{\mu \nu} 
:= (1/2)(F^I_{\mu \nu} \mp i \tilde F^I_{\mu \nu})$ 
with $F^I_{\mu \nu} = 2 \ \partial_{[\mu} A^I_{\nu]}$ 
and $\tilde F^I_{\mu \nu} 
= (1/2) \epsilon_{\mu \nu \rho \sigma} F^{I \rho \sigma}$. 
The covariant derivative $D^{(+)}_\mu$ and 
the curvature ${R^{(+)ij}}_{\mu \nu}$ are 
\ba
\A \A D^{(+)}_{\mu} := \partial_{\mu} 
      + {i \over 2} A^{(+)}_{ij \mu} 
      S^{ij}, \nonu
\A \A {R^{(+)ij}}_{\mu \nu} 
      := 2(\partial_{[\mu} {A^{(+)ij}}_{\nu]} 
      + {A^{(+)i}}_{k [\mu} {A^{(+)kj}}_{\nu]}), 
\label{D+}
\ea
while $\hat F^I_{\mu \nu}$ in the chiral Lagrangian 
density (\ref{LN3}) is defined as 
\be
\hat F^I_{\mu \nu} 
:= F^I_{\mu \nu} - {\kappa \over \sqrt{2}} 
(\overline \psi^J_{L \mu} \psi^K_{R \nu} 
+ \overline{\tilde \psi}^J_{R \mu} 
\tilde \psi^K_{L \nu}) \epsilon^{IJK}. 
\ee
In Eq. (\ref{LN3}), the $A^I_{\mu}$-dependent terms 
and the four-fermion contact terms except 
for the last two contact terms correspond to those 
obtained in the usual SUGRA. 
Note that Eq. (\ref{LN3}) is reduced 
to the $N = 2$ chiral Lagrangian density 
\cite{TS1}, if we put the condition 
\be
(\chi, \tilde \chi) = 0, \ \ \ 
(\psi^3_{\mu}, \tilde \psi^3_{\mu}) = 0, \ \ \ 
F^1_{\mu \nu} = 0 = F^2_{\mu \nu}, 
\label{reduce}
\ee
in Eq. (\ref{LN3}), and if the $e^i_{\mu}$, 
($\psi^I_{\mu}, \tilde \psi^I_{\mu}$) 
($I = 1, 2$) and $F_{\mu \nu} (:= F^3_{\mu \nu})$ 
are taken as the field variables. 

We show that the chiral Lagrangian density 
(\ref{LN3}) is invariant under the local SUSY 
transformations. 
The invariance up to the order $\kappa$ 
is confirmed by means of the following 
right- and left-handed SUSY transformations 
in the first-order formalism, 
and entirely determines the form of the chiral 
Lagrangian density. 
In particular, the last two contact terms 
in Eq. (\ref{LN3}), which are added 
by the prescription (e) of constructing 
the chiral Lagrangian density, are the non-minimal 
terms required from the invariance of order $\kappa$ 
under the right-handed SUSY transformations 
generated by $\alpha^I$ given by 
\ba
\A \A \delta_R e^i_{\mu} 
               = i \kappa \ \overline \alpha^I_L 
               \gamma^i \tilde \psi^I_{L \mu}, 
               \ \ \ \ \ 
      \delta_R A^I_{\mu} 
               = \sqrt{2} \ \epsilon^{IJK} 
               \overline \alpha^J_L \psi^K_{R \mu} 
               + \overline \alpha^I_L \gamma_\mu 
               \tilde \chi_L, 
\nonu
\A \A \delta_R \psi^I_{R \mu} 
               = {2 \over \kappa} D^{(+)}_{\mu} 
               \alpha^I_R 
               - {i \over 2} \kappa \ (\overline{\tilde \chi}_R 
               \gamma^\lambda \chi_R) \gamma_\lambda 
               \gamma_\mu \alpha^I_R, 
\nonu
\A \A \delta_R \tilde \psi^I_{L \mu} 
               = - {1 \over \sqrt{2}} \ \epsilon^{IJK} 
               \overline F{^{(-)J}}_{\rho \sigma} 
               S^{\rho \sigma} \gamma_{\mu} \alpha^K_R 
               - {i \over \sqrt{2}} \kappa 
               \ \epsilon^{IJK} 
               (\overline \psi^J_{L \mu} \gamma_\nu 
               \tilde \chi_L) \gamma^\nu \alpha^K_R, 
\nonu
\A \A \delta_R \chi_R 
               = \overline F{^{(+)I}}_{\mu \nu} 
               S^{\mu \nu} \alpha^I_R, 
               \ \ \ \ \ \ \ 
      \delta_R \tilde \chi_L = 0, 
\label{RSUSY}
\ea
and under the left-handed SUSY transformations generated 
by $\tilde \alpha^I$ given by 
\ba
\A \A \delta_L e^i_{\mu} 
               = i \kappa \ \overline{\tilde \alpha}^I_R 
               \gamma^i \psi^I_{R \mu}, 
               \ \ \ \ \ 
      \delta_L A^I_{\mu} 
               = \sqrt{2} \ \epsilon^{IJK} 
               \overline{\tilde \alpha}^J_R 
               \tilde \psi^K_{L \mu} 
               + \overline{\tilde \alpha}^I_R 
               \gamma_\mu \chi_R, 
\nonu
\A \A \delta_L \tilde \psi^I_{L \mu} 
               = {2 \over \kappa} D^{(-)}_{\mu} 
               \tilde \alpha^I_L 
               + {i \over 2} \kappa \ (\overline{\tilde \chi}_R 
               \gamma^\lambda \chi_R) \gamma_\lambda 
               \gamma_\mu \tilde \alpha^I_L, 
\nonu
\A \A \delta_L \psi^I_{R \mu} 
               = - {1 \over \sqrt{2}} \ \epsilon^{IJK} 
               \overline F{^{(+)J}}_{\rho \sigma} 
               S^{\rho \sigma} \gamma_{\mu} 
               \tilde \alpha^K_L 
               - {i \over \sqrt{2}} \kappa 
               \ \epsilon^{IJK} 
               (\overline{\tilde \psi}^J_{R \mu} 
               \gamma_\nu \chi_R) \gamma^\nu 
               \tilde \alpha^K_L, 
\nonu
\A \A \delta_L \chi_R = 0, \ \ \ \ \ \ \ 
      \delta_L \tilde \chi_L 
               = \overline F{^{(-)I}}_{\mu \nu} 
               S^{\mu \nu} \tilde \alpha^I_L, 
\label{LSUSY}
\ea
where $\overline F^I_{\mu \nu}$ 
in Eqs. (\ref{RSUSY}) and (\ref{LSUSY}) is defined as 
\be
\overline F^I_{\mu \nu} := 
\hat F^I_{\mu \nu} 
- \kappa (\overline \psi^I_{L \mu} 
\gamma_\nu \tilde \chi_L 
+ \overline{\tilde \psi}^I_{R \mu} 
\gamma_\nu \chi_R). 
\ee
In addition, we choose the right- and left-handed SUSY 
transformations of $A^{(+)}_{ij \mu}$ as \cite{TS3} 
\ba
\A \A \delta_R A^{(+)}_{ij \mu} = 0, 
\nonu
\A \A \delta_L A^{(+)}_{ij \mu} 
      = {\rm self\! \! -\! \! dual\ part\ of\ } \{ - \kappa 
      (B_{\mu ij} - e_{\mu [i} B{^m}_{\mid m \mid j]}) \}, 
\label{ASUSY}
\ea
respectively, with 
\be
B^{\lambda \mu \nu} 
:= \epsilon^{\mu \nu \rho \sigma} 
   \overline{\tilde \alpha}^I_R \gamma^{\lambda} 
   D^{(+)}_{\rho} \psi^I_{R \sigma}. 
\ee
Although the $A^{(-)}_{ij \mu}$ 
appears at $\delta_L \tilde \psi^I_{L \mu}$ 
in the left-handed SUSY transformations 
of Eq. (\ref{LSUSY}), it is not an independent 
variable but a quantity given by $e^i_\mu$, \ 
$(\psi^I_{\mu}, \tilde \psi^I_{\mu})$, 
and $(\chi, \tilde \chi)$; namely, 
the $A^{(-)}_{ij \mu}$ is fixed as the sum 
of the antiself-dual part of the Ricci rotation 
coefficients $A_{ij \mu}(e)$ and that of $K_{ij \mu}$ 
given by 
\be
K_{ij \mu} = {i \over 2} \kappa^2 
             (\overline{\tilde \psi}^I_{R [i} 
             \gamma_{\mid \mu \mid} \psi^I_{R j]} 
             + \overline{\tilde \psi}^I_{R [i} 
             \gamma_{\mid j \mid} \psi^I_{R \mu]} 
             - \overline{\tilde \psi}^I_{R [j} 
             \gamma_{\mid i \mid} \psi^I_{R \mu]}) 
             + {\kappa^2 \over 4} \epsilon_{ij \mu \nu} 
             \overline{\tilde \chi}_R \gamma^\nu \chi_R. 
\ee

At order $\kappa^2$ and $\kappa^3$, 
on the other hand, the transformations 
(\ref{ASUSY}) should be corrected to recover the SUSY 
invariance in the first-order formalism. 
However, this task will be complicated as is expected 
from the usual SUGRA \cite{Fr}, and therefore we turn to 
the second-order formalism in order to minimize 
complication. Then it can be shown by a straightforward 
calculation that the chiral Lagrangian density 
(\ref{LN3}) is invariant under the SUSY transformations 
of Eqs. (\ref{RSUSY}) and (\ref{LSUSY}), 
provided that the $A^{(+)}_{ij \mu}$ is fixed as 
\be
A^{(+)}_{ij \mu} = A^{(+)}_{ij \mu}(e) 
+ K^{(+)}_{ij \mu} 
\label{A+}
\ee
by solving the equation 
$\delta {\cal L}^{(+)}_{N = 3}/\delta A^{(+)}_{ij \mu} 
= 0$ with respect to $A^{(+)}_{ij \mu}$. 

Here we also note that the last two four-fermion 
contact terms in Eq. (\ref{LN3}), which do not appear 
in the first-order Lagrangian density \cite{Fr,FSZ,Nieu} 
of the usual SUGRA, are necessary to reproduce 
the second-order Lagrangian density 
of the usual $N = 3$ SUGRA, when the reality condition, 
\be
\overline{e^i_{\mu}} = e^i_{\mu}, \ \ \ 
\tilde \psi^I_{\mu} = \psi^I_{\mu}, \ \ \ 
\tilde \chi = \chi 
{\rm\ \ and\ \ } 
\overline{A^I_{\mu}} = A^I_{\mu}, 
\label{reality}
\ee
is imposed. 
\footnote{
The bars of $e^i_{\mu}$ and $A^I_{\mu}$ 
in Eq. (\ref{reality}) mean the complex conjugate.}
Indeed, if we use the solution (\ref{A+}) 
in the first three terms in Eq. (\ref{LN3}), 
then these terms give rise to a number of four-fermion 
contact terms, which involve the following terms 
\ba
{i \over {8 \kappa^2}} 
e \ \epsilon^{\mu \nu \rho \sigma} 
T_{\lambda \mu \nu} T{^{\lambda}}_{\rho \sigma} 
= \A \A - {i \over{16}} \kappa^2 
  e \ \epsilon^{\mu \nu \rho \sigma} 
  (\overline{\tilde \psi}^J_{R \mu} 
  \gamma_{\lambda} \psi^L_{R \nu}) 
  \overline{\tilde \psi}^K_{R \rho} \gamma^{\lambda} 
  \psi^M_{R \sigma} \epsilon^{IJK} \epsilon^{ILM} 
\nonu
  \A \A - {\kappa^2 \over 2} e \ 
  (\overline{\tilde \psi}^I_{R \mu} 
  \gamma^{[\mu} \psi^I_{R \nu}) \ 
  \overline{\tilde \chi}_R \gamma^{\nu]} \chi_R, 
\label{4-Fermi}
\ea
where the torsion tensor is defined by 
$T{^i}_{\mu \nu} = - 2 D_{[\mu} e^i_{\nu]}$ 
with $D_{\mu} e^i_{\nu} = \partial_{\mu} e^i_{\nu} 
+ A{^i}_{j \mu} e^j_{\nu}$. 
The second last term in Eq. (\ref{LN3}), 
on the other hand, can be rewritten as 
\ba
\A \A {i \over 8} \kappa^2 
      e \ \epsilon^{\mu \nu \rho \sigma} 
      (\overline \psi^J_{L \mu} \psi^K_{R \nu}) 
      \overline{\tilde \psi}^L_{R \rho} 
      \tilde \psi^M_{L \sigma} 
      \epsilon^{IJK} \epsilon^{ILM} 
\nonu
\A \A \ \ \ \ = {i \over{16}} \kappa^2 
      e \ \epsilon^{\mu \nu \rho \sigma} 
      (\overline{\tilde \psi}^J_{R \mu} 
      \gamma_{\lambda} \psi^L_{R \nu}) 
      \overline{\tilde \psi}^K_{R \rho} 
      \gamma^{\lambda} \psi^M_{R \sigma} 
      \epsilon^{IJK} \epsilon^{ILM} 
\label{Fierz}
\ea
by using a Fierz transformation, and the sum of 
the last two terms in Eq. (\ref{LN3}) 
exactly cancels out the terms of Eq. (\ref{4-Fermi}). 
When the reality condition (\ref{reality}) 
is imposed in the second-order formalism, 
the last two terms in Eq. (\ref{reality}) 
and the terms of Eq. (\ref{4-Fermi}) are purely 
imaginary up to boundary terms, 
but they cancel with each other. 
Therefore the ${\cal L}^{(+)}_{N = 3}[{\rm second\ order}]$ 
of $N = 3$ chiral SUGRA with the reality condition 
(\ref{reality}) is reduced to that of the usual one 
up to imaginary boundary terms; 
namely, we have 
\footnote{
The imaginary boundary terms in Eq. (\ref{LN3S}) 
correspond to the Chern-Simons type boundary terms 
\cite{Mac,MMT} as appeared in Eq. (\ref{LN2}) 
evaluated in the second-order formalism.} 
\ba
{\cal L}^{(+)}_{N = 3}[{\rm second\ order}] 
= \A \A {\cal L}_{N = 3 {\rm\ usual\ SUGRA}} 
[{\rm second\ order}] \nonu
\A \A + {1 \over 8} \partial_{\mu} 
(e \ \epsilon^{\mu \nu \rho \sigma} 
\overline \psi^I_{\rho} \gamma_{\nu} 
\psi^I_{\sigma} 
+ i e \ \overline \chi \gamma_5 \gamma^{\mu} \chi). 
\label{LN3S}
\ea
This second-order Lagrangian density (\ref{LN3S}) 
is invariant under the right- and left-handed SUSY 
transformations of Eqs. (\ref{RSUSY}) and (\ref{LSUSY}), 
which are now complex conjugate of each other 
in the second-order formalism 
under the reality condition (\ref{reality}). 

Let us explain how to gauge \cite{FrDa} 
the global $O(3)$ invariance of the chiral Lagrangian 
density (\ref{LN3}). Firstly we introduce 
a minimal coupling of $\psi^I_{R \mu}$ with $A^I_{\mu}$, 
which automatically requires a spin-3/2 mass-like 
term and a cosmological term in order to ensure 
the SUSY invariance of the Lagrangian, 
and we also replace the Abelian field 
strength $F^I_{\mu \nu}$ with the non-Abelian one, 
\be
F'^I_{\mu \nu} := F^I_{\mu \nu} 
+ \lambda \epsilon^{IJK} A^J_{\mu} A^K_{\nu} 
\ee
with the gauge coupling constant $\lambda$: 
The three terms added to Eq. (\ref{LN3}) 
in order to gauge the $O(3)$ invariance 
are then written as 
\ba
{\cal L}_{{\rm cosm}} 
= \A \A - \lambda e \ \epsilon^{\mu \nu \rho \sigma} 
        \overline{\tilde \psi}^I_{R \mu} 
        \gamma_\rho \psi^K_{R \nu} A^J_\sigma 
        \epsilon^{IJK} 
\nonu
\A \A - \sqrt{2} i \kappa^{-1} \lambda e 
        (\overline{\tilde \psi}^I_{R \mu} 
        S^{\mu \nu} \tilde \psi^I_{L \nu} 
        + \overline \psi^I_{L \mu} 
        S^{\mu \nu} \psi^I_{R \nu}) 
\nonu
\A \A - \Lambda \kappa^{-2} e, 
\ea
where the cosmological constant $\Lambda$ is related 
to $\lambda$ as $\Lambda = - 6 \kappa^{-2} \lambda^2$.

\newsection{Duality invariance in $N = 3$ chiral SUGRA}

In the extended usual SUGRA without gauging 
the global O($N$) invariance, the field equation 
for vector fields is invariant under duality 
transformations \cite{FSZ2,GZ,Tanii} 
which generalize those of the free Maxwell equations, 
while the Lagrangian changes its form in a specific way 
under these transformations. 

As we have already noted, the chiral Lagrangian density 
(\ref{LN3}) possesses global O(3) invariance. 
We show in this section that the field equation 
for vector fields derived from the nongauged, 
chiral Lagrangian density (\ref{LN3}) is invariant 
under duality transformations. 
The field equation for $A^I_{\mu}$ can be written as 
\be
\partial_{\mu} (e \tilde G^{I \mu \nu}) = 0, 
\label{eqG}
\ee
where the $\tilde G^{I \mu \nu}$ 
are defined by 
\be
\tilde G^{I \mu \nu} 
:= {2 \over e} \ 
{{\partial {\cal L}^{(+)}_{N = 3}} \over 
{\partial F^I_{\mu \nu}}} 
= - F^{I \mu \nu} + \kappa (H^{(+)I \mu \nu} 
+ I^{(-)I \mu \nu}) 
\ee
with $H^{(+)I \mu \nu}$ and $I^{(-)I \mu \nu}$ 
being given by 
\ba
\A \A H^{(+)I \mu \nu} = \sqrt{2} \ \epsilon^{IJK} 
(\overline{\tilde \psi}{^J_R}^{\mu} 
\tilde \psi{^K_L}^{\nu})^{(+)} 
- i \ \overline \psi^I_{L \lambda} S^{\mu \nu} 
\gamma^{\lambda} \tilde \chi_L, 
\nonu
\A \A I^{(-)I \mu \nu} = \sqrt{2} \ \epsilon^{IJK} 
(\overline \psi{^J_L}^{\mu} \psi{^K_R}^{\nu})^{(-)} 
- i \ \overline{\tilde \psi}^I_{R \lambda} S^{\mu \nu} 
\gamma^{\lambda} \chi_R. 
\label{H+I-}
\ea
In addition, the $\tilde F^I_{\mu \nu}$ satisfies 
the Bianchi identity 
\be
\partial_{\mu} (e \tilde F^{I \mu \nu}) = 0. 
\label{eqF}
\ee
Equations (\ref{eqG}) and (\ref{eqF}) 
are invariant under the following (global) 
duality transformations, 
\ba
\A \A \delta e^i_{\mu} = 0, \ \ \ 
\delta \psi^I_{R \mu} 
= - i \Lambda^{IJ} \psi^J_{R \mu}, \ \ \ 
\delta \tilde \psi^I_{L \mu} 
= i \Lambda^{IJ} \tilde \psi^J_{L \mu}, 
\nonu
\A \A \delta \chi_R = 0, \ \ \ \delta \tilde \chi_L = 0, 
\nonu
\A \A \delta \pmatrix{ F^{I \mu \nu} \cr
                 G^{I \mu \nu}}
     = \pmatrix{ 0 & \Lambda^{IJ} \cr
                 - \Lambda^{IJ} & 0}
       \pmatrix{ F^{J \mu \nu} \cr
                 G^{J \mu \nu}} 
\label{dualtr1}
\ea
with the constant parameters $\Lambda^{IJ}$ 
which are assumed to be complex, 
symmetric ($\Lambda^{IJ} = \Lambda^{JI}$) 
and traceless ($\Lambda^{II} = 0$): 
When the reality condition (\ref{reality}) 
is imposed, however, $\Lambda^{IJ}$ are supposed 
to be real. The transformations of Eq. (\ref{dualtr1}) 
can be rewritten in terms of 
the bases $(H^{(+)I \mu \nu}, I^{(-)I \mu \nu})$ 
and $(F^{I \mu \nu} + i G^{I \mu \nu}, 
F^{I \mu \nu} - i G^{I \mu \nu})$ as 
\ba
\delta \pmatrix{
H^{(+)I \mu \nu} \cr
I^{(-)I \mu \nu}}
= \A \A - i \pmatrix{
  \Lambda^{IJ} & 0 \cr
  0 & - \Lambda^{IJ}}
  \pmatrix{
  H^{(+)J \mu \nu} \cr 
  I^{(-)J \mu \nu}}, 
\nonu
\delta \pmatrix{
F^{I \mu \nu} + i G^{I \mu \nu} \cr
F^{I \mu \nu} - i G^{I \mu \nu}}
= \A \A - i \pmatrix{
  \Lambda^{IJ} & 0 \cr
  0 & - \Lambda^{IJ}}
  \pmatrix{
  F^{J \mu \nu} + i G^{J \mu \nu} \cr
  F^{J \mu \nu} - i G^{J \mu \nu}}. 
\label{dualtr2}
\ea
Since the transformations of Eq. (\ref{dualtr2}) 
combined with the O(3) transformations 
becomes the SU(3) group \cite{FSZ2}, 
the duality symmetry based on Eq. (\ref{dualtr2}) 
is an example for ``compact" duality symmetries 
\cite{GZ}. Also, the transformations (\ref{dualtr2}) 
reduce to (global) U(1) transformations 
in $N = 2$ chiral SUGRA 
if Eq. (\ref{reduce}) and $\Lambda^{11} = \Lambda^{22} 
= -(1/2) \Lambda^{33}$ are satisfied. 

The $N = 3$ chiral Lagrangian density (\ref{LN3}) 
is expressed by using 
$(F^{I \mu \nu}, G^{I \mu \nu})$ 
and $(H^{(+)I \mu \nu}, I^{(-)I \mu \nu})$ as 
\ba
{\cal L}^{(+)}_{N = 3} 
= \A \A {e \over 4} (F^I_{\mu \nu} 
        \tilde G^{I \mu \nu}) 
\nonu
\A \A + {e \over 8} 
      \{ (F^{(+)I}_{\mu \nu} - i G^{(+)I}_{\mu \nu}) 
      H^{(+)I \mu \nu} 
      + (F^{(-)I}_{\mu \nu} + i G^{(-)I}_{\mu \nu}) 
      I^{(-)I \mu \nu} \} 
\nonu
\A \A + [A^I_{\mu}{\rm -independent\ terms}]. 
\label{LN3dual}
\ea
The second line is obviously invariant under 
the transformation of Eq. (\ref{dualtr2}), 
and the invariance of the $A^I_{\mu}$-independent 
terms can also be confirmed by using Eq. (\ref{dualtr1}). 
Therefore, the ${\cal L}^{(+)}_{N = 3}$ transforms 
under Eq. (\ref{dualtr2}) in a definite way as 
\ba
\delta {\cal L}^{(+)}_{N = 3} 
\A \A = \delta \left( {e \over 4} 
        F^I_{\mu \nu} \tilde G^{I \mu \nu} \right) 
\nonu
\A \A = - {e \over 4} (F^I_{\mu \nu} 
      \Lambda^{IJ} \tilde F^{J \mu \nu} 
      - \tilde G^I_{\mu \nu} 
      \Lambda^{IJ} G^{J \mu \nu}), 
\ea
which is same as that of the usual SUGRA 
except that the parameters $\Lambda^{IJ}$ 
are now complex.

\newsection{Construction of the higher $N$ 
chiral Lagrangian density} 

In this section let us first construct the $N = 4$ 
chiral Lagrangian density by applying 
the prescription (a)-(e) explained in Introduction. 
In the usual $N = 4$ SUGRA, the field contents 
are a tetrad field, four Rarita-Schwinger 
fields, six vector fields, four spin-1/2 fields 
and two (real) scalar fields. 
Since there are scalar fields in the theory, 
the duality symmetry group becomes ``non-compact" 
\cite{GZ} in contrast with the $N = 2, 3$ theories. 
Indeed, the duality symmetry group of the usual 
$N = 4$ SUGRA is SU(4) $\times$ SU(1,1) \cite{CSF}, 
and the two scalar fields are described 
by the SU(1,1)/U(1) non-linear sigma model 
\cite{GZ,Tanii}. 

In $N = 4$ chiral SUGRA, 
we introduce at first the (complex) tetrad $e^i_{\mu}$, 
two independent sets of Rarita-Schwinger 
fields ($\psi^I_{\mu}, \tilde \psi^I_{\mu}$) 
($I = 1, 2, 3, 4$), two independent sets 
of spin-1/2 fields ($\chi^I$, $\tilde \chi^I$), 
(complex) vector fields $A^{IJ}_{\mu} 
(= - A^{JI}_{\mu})$ and complex scalar fields 
\footnote{
In $N = 4$ chiral SUGRA, we define complex scalar 
fields as ${}^{\pm} \phi_1 = p \pm i q$ 
and ${}^{\pm} \phi_2 = r \pm i s$ 
with $p, q, r$ and $s$ being assumed to be complex, 
respectively. 
The reality condition for these scalar fields 
will be taken as ($\overline{{}^- \phi{_1}}, \ 
\overline{{}^- \phi{_2}}$) 
= (${}^+ \phi_1, {}^+ \phi_2$).} 
(${}^{\pm} \phi_1$, ${}^{\pm} \phi_2$) 
as the field variables. 
The self-dual connection $A^{(+)}_{ij \mu}$ is also 
treated as one of the independent variables 
in the first-order formalism. 
If we apply the prescription of constructing 
the chiral Lagrangian density from the usual $N = 4$ SUGRA 
as in the case of $N = 3$, then the obtained $N = 4$ 
chiral Lagrangian density in first-order form 
can be written schematically as 
\ba
{\cal L}^{(+)}_{N = 4} 
\A = \A - {i \over {2 \kappa^2}} 
        e \ \epsilon^{\mu \nu \rho \sigma} 
        e^i_{\mu} e^j_{\nu} R^{(+)}_{ij \rho \sigma} 
      - e \ \epsilon^{\mu \nu \rho \sigma} 
        \overline{\tilde \psi}^I_{R \mu} \gamma_\rho 
        D^{(+)}_\sigma \psi^I_{R \nu} 
\nonu
\A \A + i e \ \overline{\tilde \chi}^I_R 
        \gamma^\mu 
        D^{(+)}_\mu \chi^I_R 
\nonu
\A \A + {\cal L}_{N = 4}[{\rm scalar\ kinetic\ term} 
        + A^{IJ}_{\mu}{\rm -dependent\ terms}] 
\nonu
\A \A + [A^{IJ}_{\mu}{\rm -independent\ terms}]
\nonu
\A \A + {i \over {16}} \kappa^2 
        e \ \epsilon^{\mu \nu \rho \sigma} 
        (\overline \psi^K_{L \mu} \psi^L_{R \nu}) 
        \overline{\tilde \psi}^M_{R \rho} 
        \tilde \psi^N_{L \sigma} 
        \epsilon^{IJKL} \epsilon^{IJMN} 
\nonu
\A \A + {\kappa^2 \over 2} e \ 
        (\overline{\tilde \psi}^I_{R \mu} 
        \gamma^{[\mu} \psi^I_{R \nu}) \ 
        \overline{\tilde \chi}^J_R \gamma^{\nu]} 
        \chi^J_R 
\label{LN4}
\ea
with 
\ba
\A \A {\cal L}_{N = 4}[{\rm scalar\ kinetic\ term} 
+ A^{IJ}_{\mu}{\rm -dependent\ terms}] 
\nonu
\A \A \hspace{2cm} = {1 \over 2} \ 
      {{\partial_{\mu} {}^+ z \ 
      \partial^{\mu} {}^- z} \over 
      {(1 - {}^+ z {}^- z)}} 
      - {e \over 4} F^{IJ}_{\mu \nu} 
      K_{IJ,KL} F^{KL \mu \nu} 
\nonu
\A \A \hspace{2.5cm} + [F^{IJ}_{\mu \nu} 
      {\rm -proportional\ terms}]. 
\label{LN4sc}
\ea
The first term in Eq. (\ref{LN4sc}) is the kinetic term 
of the scalar fields corresponding to 
the SU(1,1)/U(1) non-linear sigma model, 
and ${}^{\pm} z$ in this term is defined as 
U(1) invariant variable constructed from 
(${}^{\pm} \phi_1$, ${}^{\pm} \phi_2$), i.e., 
\be
{}^{\pm} z := {}^{\pm} \phi_2 
({}^{\pm} \phi{_1})^{-1}. 
\ee
The function $K_{IJ,KL} (= K_{KL,IJ})$ 
in the second term of Eq. (\ref{LN4sc}), 
on the other hand, is given by 
\be
K_{IJ,KL} = {{1 + {}^- z^2} \over {1 - {}^- z^2}} \ 
\delta_{I[K} \delta_{\mid J \mid L]} 
- {{2 {}^- z} \over {1 - {}^- z^2}} {1 \over 2} \ 
\epsilon_{IJKL}, 
\ee
which is determined from a specific transformation 
property of $K_{IJ,KL}$ \cite{GZ,Tanii} 
under the duality transformations. 
Eq. (\ref{LN4sc}) and the $A^{IJ}_{\mu}$-independent 
terms in Eq. (\ref{LN4}) correspond to those obtained 
in the usual $N = 4$ SUGRA \cite{CSF,CS}. 
In order to prove the SUSY invariance 
of the chiral Lagrangian density (\ref{LN4}) 
under right- and left-handed SUSY transformations, 
we will need a straightforward calculation. 

The last two four-fermion contact terms 
in Eq. (\ref{LN4}) has the same role as in $N = 2, 3$ 
chiral SUGRA; 
namely, those terms ensure the first-order SUSY 
invariance and are also necessary to reproduce 
the Lagrangian density of the usual $N = 4$ SUGRA, 
when the reality condition, 
\be
\overline{e^i_{\mu}} = e^i_{\mu}, \ \ \ 
\tilde \psi^I_{\mu} = \psi^I_{\mu}, \ \ \ 
\tilde \chi^I = \chi^I, \ \ \ 
\overline{A^{IJ}_{\mu}} = A^{IJ}_{\mu} 
{\rm\ \ and\ \ } 
(\overline{{}^- \phi{_1}}, \ 
\overline{{}^- \phi{_2}}) = 
({}^+ \phi_1, {}^+ \phi_2), 
\label{reality2}
\ee
is imposed. 
Indeed, the last two terms in Eq. (\ref{LN4}) 
exactly cancel with two of terms obtained 
in the second-order formalism by solving the equation 
$\delta {\cal L}^{(+)}_{N = 4}/\delta A^{(+)}_{ij \mu} 
= 0$ with respect to $A^{(+)}_{ij \mu}$. 
Then the ${\cal L}^{(+)}_{N = 4}[{\rm second\ order}]$ 
of $N = 4$ chiral SUGRA with the reality condition 
(\ref{reality2}) is reduced to that of the usual one 
up to imaginary boundary terms as 
\ba
{\cal L}^{(+)}_{N = 4}[{\rm second\ order}] 
= \A \A {\cal L}_{N = 4 {\rm\ usual\ SUGRA}} 
[{\rm second\ order}] \nonu
\A \A + {1 \over 8} \partial_{\mu} 
(e \ \epsilon^{\mu \nu \rho \sigma} 
\overline \psi^I_{\rho} \gamma_{\nu} 
\psi^I_{\sigma} 
+ i e \ \overline \chi^I \gamma_5 
\gamma^{\mu} \chi^I). 
\label{LN4S}
\ea
This second-order Lagrangian density (\ref{LN4S}) 
is invariant under the SUSY transformations 
of the usual $N = 4$ SUGRA. 

Next we discuss the construction of the $N = 8$ 
chiral Lagrangian density also by applying 
the prescription (a)-(e) explained in Introduction. 
Here we note the characteristic features 
of the chiral Lagrangian density constructed so far. 
The $N = 3$ and $4$ chiral Lagrangian densities 
of Eqs. (\ref{LN3}) and (\ref{LN4}) have different 
forms from those of the usual SUGRA, in particular, 
with respect to the following points: 
Firstly, only the gravitational and spinor 
(spin-3/2 and -1/2) kinetic terms in the chiral 
Lagrangian density are written in terms of 
the self-dual connection $A^{(+)}_{ij \mu}$. 
Then the appropriate four-fermion contact terms, 
which are required from the invariance 
under first-order SUSY transformations 
at order $\kappa$ at least up to $N = 3$ chiral SUGRA, 
are added in the chiral Lagrangian density 
by the prescription (e). 
In view of these points, the prescription of 
constructing the chiral Lagrangian density 
is easily extended to $N = 8$ SUGRA \cite{CJ,dWN}. 

The field contents of the usual $N = 8$ SUGRA 
are a tetrad field, eight Rarita-Schwinger 
fields, 28 vector fields, 56 spin-1/2 fields 
and 35 complex scalar fields. 
If we introduce at first the (complex) tetrad $e^i_{\mu}$, 
two independent sets of Rarita-Schwinger 
fields ($\psi^I_{\mu}, \tilde \psi^I_{\mu}$) 
($I = 1, \dots, 8$), two independent sets 
of spin-1/2 fields ($\chi^{IJK}$, $\tilde \chi^{IJK}$), 
\footnote{
The $\chi^{IJK}$ denotes totally antisymmetric spinor, 
i.e., $\chi^{IJK} = \chi^{[IJK]}$.}
then the gravitational and spinor kinetic terms 
written by the $A^{(+)}_{ij \mu}$ 
in the $N = 8$ chiral Lagrangian density 
${\cal L}^{(+)}_{N = 8}$ are written as 
\ba
\A \A {\cal L}^{(+)}_{N = 8}[{\rm gravitational\ and\ 
spinor\ kinetic\ terms}] 
\nonu
\A \A \hspace{2cm} 
= - {i \over {2 \kappa^2}} 
  e \ \epsilon^{\mu \nu \rho \sigma} 
  e^i_{\mu} e^j_{\nu} R^{(+)}_{ij \rho \sigma} 
  - e \ \epsilon^{\mu \nu \rho \sigma} 
  \overline{\tilde \psi}^I_{R \mu} \gamma_\rho 
  D^{(+)}_\sigma \psi^I_{R \nu} 
\nonu
\A \A \hspace{2.5cm} + {i \over 6} 
  e \ \overline{\tilde \chi}^{IJK}_R 
  \gamma^\mu D^{(+)}_\mu \chi^{IJK}_R. 
\label{LN8k}
\ea
The four-fermion contact terms added 
in ${\cal L}^{(+)}_{N = 8}$ by means of 
the prescription (e), on the other hand, 
are chosen as 
\ba
\A \A {\cal L}^{(+)}_{N = 8}
      [{\rm contact\ terms\ by\ the\ prescription\ (e)}] 
\nonu
\A \A \ \ \ \ = {i \over {8 \cdot 6!}} \kappa^2 
        e \ \epsilon^{\mu \nu \rho \sigma} 
        (\overline \psi^P_{L \mu} \psi^Q_{R \nu}) 
        \overline{\tilde \psi}^R_{R \rho} 
        \tilde \psi^S_{L \sigma} 
        \epsilon^{IJKLMNPQ} \epsilon^{IJKLMNRS} 
\nonu
\A \A \ \ \ \ \ \ \ + {\kappa^2 \over {12}} e \ 
        (\overline{\tilde \psi}^I_{R \mu} 
        \gamma^{[\mu} \psi^I_{R \nu}) \ 
        \overline{\tilde \chi}^{JMN}_R \gamma^{\nu]} 
        \chi^{JMN}_R, 
\label{LN8f}
\ea
which will also be required from the invariance 
under first-order SUSY transformations 
at order $\kappa$. The terms other than 
Eqs. (\ref{LN8k}) and (\ref{LN8f}) correspond 
to those obtained in the usual $N = 8$ SUGRA 
\cite{CJ,dWN}. In order to prove the SUSY invariance 
of ${\cal L}^{(+)}_{N = 8}$ under right- and 
left-handed SUSY transformations, 
we will also need a straightforward calculation. 

By means of the four-fermion contact terms of 
Eq. (\ref{LN8f}), 
the ${\cal L}^{(+)}_{N = 8}[{\rm second\ order}]$ 
of $N = 8$ chiral SUGRA with the reality condition 
is also reduced to that of the usual one 
up to imaginary boundary terms as 
\ba
{\cal L}^{(+)}_{N = 8}[{\rm second\ order}] 
= \A \A {\cal L}_{N = 8 {\rm\ usual\ SUGRA}} 
[{\rm second\ order}] \nonu
\A \A + {1 \over 8} \partial_{\mu} 
\left( e \ \epsilon^{\mu \nu \rho \sigma} 
\overline \psi^I_{\rho} \gamma_{\nu} 
\psi^I_{\sigma} 
+ {i \over 6} e \ \overline \chi^{IJK} 
\gamma_5 \gamma^{\mu} \chi^{IJK} \right), 
\label{LN8S}
\ea
which is invariant under the SUSY transformations 
of the usual $N = 8$ SUGRA. 
The imaginary boundary terms in Eq. (\ref{LN8S}) 
are same as those of the $N = 3, 4$ chiral SUGRA.

\newsection{Conclusion}

In this paper we obtained $N = 3$ chiral SUGRA 
compatible with the reality condition 
by applying the prescription of constructing 
the chiral Lagrangian density from the usual 
$N = 3$ SUGRA. The $N = 3$ chiral Lagrangian density 
in first-order form of Eq. (\ref{LN3}) 
was determined so that it reproduces 
the ${\cal L}^{(+)}_{N = 3}[{\rm second\ order}]$ 
of Eq. (\ref{LN3S}) by adding the appropriate 
four-fermion contact terms, 
and showed that those four-fermion contact terms 
added in Eq. (\ref{LN3}) are the non-minimal terms 
required from the invariance under the first-order 
SUSY transformations at order $\kappa$. 
We also showed that the field equation 
for the vector fields derived from Eq. (\ref{LN3}) 
is invariant under the (compact) duality transformations. 

Furthermore, we constructed the $N = 4$ chiral Lagrangian 
density, in which the  duality symmetry 
group is (non-compact) SU(4) $\times$ SU(1,1), 
and we also discussed the construction of the $N = 8$ 
chiral Lagrangian density. 
In the higher $N$ chiral Lagrangian density 
we added appropriate four-fermion contact terms 
as in the case of $N = 3$, 
which will be required from the invariance 
under the first-order SUSY transformations 
at order $\kappa$. We will need a straightforward 
calculation in order to prove the SUSY invariance 
of the higher $N$ chiral Lagrangian density 
under right- and left-handed SUSY transformations. 

Finally we briefly discuss the polynomiality 
of constraints in the canonical formulation 
of the chiral SUGRA. There appear, in the chiral SUGRA, 
right- and left-handed SUSY constraints in addition to 
Gauss-law, U(1) gauge (for $N \ge 2$), vector 
and Hamiltonian constraints, which reflect 
the invariance of the chiral Lagrangian density. 
In the $N = 1$ theory \cite{Jac}, all the constraints 
are indeed written in polynomial form in terms of 
the canonical variables of the Ashtekar formulation. 
In the $N = 2$ theory \cite{KS,TS2}, although only 
the left-handed SUSY constraint 
(and the Hamiltonian constraint as stated in \cite{KS}) 
has the non-polynomial factor as in the case of 
the Einstein-Maxwell theory in the Ashtekar variable 
\cite{GP}, the rescaled left-handed SUSY constraint 
by multiplying this factor becomes polynomial. 
In the $N = 3$ theory derived from the $N = 3$ 
chiral Lagrangian density (\ref{LN3}) with the reality 
condition (\ref{reality}), 
it can be verified that both right- and left-handed 
SUSY constraints have the same non-polynomial factor 
as appears only in the left-handed SUSY constraint 
of the $N = 2$ theory. 
However, the polynomiality of these SUSY constraints 
is also recovered by multiplying this factor 
to the constraints. 
The constraint algebra of the $N = 3$ theory 
is now under investigation and will be reported elsewhere.

\vspace{1.5cm}

{\large{\bf{Acknowledgments}}} 

I am grateful to Professor T. Shirafuji 
for useful discussions and reading the manuscript. 
I am also grateful to Professor Y. Tanii 
for discussions. 
I would like to thank the members of Physics Department 
at Saitama University and Laboratory of Physics 
at Saitama Institute of Technology 
for discussions and encouragements. 
This work was supported in part 
by the High-Tech Research Center 
of Saitama Institute of Technology.



\end{document}